\documentclass[12pt,a4paper]{article}

\usepackage{amsmath,amssymb,amsfonts,bm,graphicx,subfiles,xspace,hyperref}
\usepackage{soul}

\newlength{\abstwidth}
\renewcommand{\eqref}[1]{Eq.~(\ref{#1})}

\newcommand{\pT}{\ensuremath{p_{\rm T}}\xspace}

\newcommand{\XIM}{\ensuremath{\Xi^{-}}\xspace}
\newcommand{\XIP}{\ensuremath{\overline{\Xi}^{+}}\xspace}
\newcommand{\NXIM}{\ensuremath{N_{\XIM}}\xspace}
\newcommand{\NXIP}{\ensuremath{N_{\XIP}}\xspace}

\newcommand{\KM}{\ensuremath{K^{-}}\xspace}
\newcommand{\KP}{\ensuremath{K^{+}}\xspace}
\newcommand{\NKM}{\ensuremath{N_{\KM}}\xspace}
\newcommand{\NKP}{\ensuremath{N_{\KP}}\xspace}
\newcommand{\kxi}{\ensuremath{k_{\Xi}}\xspace}
\newcommand{\kk}{\ensuremath{k_{K}}\xspace}
\newcommand{\kxik}{\ensuremath{k_{\Xi K}}\xspace}
\newcommand{\sppt}[1]{\ensuremath{\sqrt{s} = #1\,\text{TeV}}\xspace}

\renewcommand{\epsilon}{\varepsilon}

\begin{document}
\sloppy
 
\pagestyle{empty}
 
\vspace{\fill}

\begin{center}
{\Huge\bf Simplifying Strangeness Fluctuations through Balance Functions in Proton-Proton Collisions}\\[4mm]
{\Large Christian Bierlich, Peter Christiansen} \\[3mm]
{\it Division of Particle and Nuclear Physics,}\\[1mm]
{\it Department of Physics,}\\[1mm]
{\it Lund University,}\\[1mm]
{\it S\"olvegatan 14A,}\\[1mm]
{\it SE-223 62 Lund, Sweden}
\end{center}

\vspace{\fill}

\begin{center}
\begin{minipage}{\abstwidth}
{\bf Abstract}\\[2mm]

Recent measurements of event-by-event fluctuations of multistrange baryons and kaons in proton-proton collisions by ALICE have been proposed as sensitive probes to distinguish between thermal and string-based hadronization mechanisms. We demonstrate that two key observables --- the normalized net-$\Xi$ second-order cumulant and the net-$K$–net-$\Xi$ Pearson correlation coefficient --- can be re-expressed in terms of balance function integrals, thereby revealing their underlying microscopic content, and relating them to balance functions previously measured by ALICE. This allows us to show that both observables probe the same physics and primarily measure the impact of strangeness conservation on hadronization. We compare both sets of ALICE data with two contrasting models, PYTHIA and Thermal-FIST. Importantly, while Thermal-FIST can reproduce the magnitude of fluctuation observables, it fails to describe the shape of the measured balance functions. Our findings highlight the importance of balance functions as precision tools for testing hadronization models in small collision systems and propose a robust baseline for future studies exploring hadronization dynamics in QCD matter.
\end{minipage}
\end{center}

\vspace{\fill}

\phantom{dummy}

\clearpage

\pagestyle{plain}
\setcounter{page}{1}
\section{Introduction}
The discovery of strangeness enhancement in proton-proton (pp) and proton-lead (p–Pb) collisions, see Refs.~\cite{ALICE:2022wpn,Christiansen:2024bhe} for recent overviews, represents a significant development in QCD physics. Strangeness enhancement was originally predicted as a Quark-Gluon Plasma (QGP) signature in large nucleus-nucleus (AA) collisions \cite{Koch:1986ud}. Therefore, it is not surprising that the experimental results can be described well by QGP-inspired statistical thermal models~\cite{Andronic:2017pug}, such as the Thermal-FIST model~\cite{Vovchenko:2019pjl} and the core-corona based EPOS model~\cite{Werner:2007bf}. On the other hand, developments in traditional vacuum-based hadronization models, that have been very successful in describing all general features of pp collisions, have also been able to quantitatively reproduce the strangeness enhancement. For example, the extension of the Lund string model~\cite{Andersson:1979ij,Andersson:1983jt,Andersson:1983ia} by rope hadronization~\cite{Bierlich:2024odg,Bierlich:2014xba}, used in the PYTHIA Monte Carlo event generator~\cite{Bierlich:2022pfr}, as well as novel cluster hadronization schemes~\cite{Gieseke:2017clv} used in HERWIG~\cite{Bellm:2015jjp}. This has lead to the situation where very different physics mechanisms can explain the strangeness enhancement and a lot of work is now ongoing to resolve which mechanism, if any of the proposed ones, is realized in nature. \\
Recently, event-by-event fluctuations of multistrange baryons, $\Xi^+$ and $\Xi^-$, and their correlation with net-kaon number quantified by the Pearson correlation coefficient, have been measured in Ref.~\cite{ALICE:2024rnr} and proposed as a discriminator between different types of mechanism. The study in Ref.~\cite{ALICE:2024rnr} observes a large discrepancy between the PYTHIA (string) predictions and data.

A big challenge in understanding the significance of this discrepancy, is the lack of understanding of the coupling between the measured quantities and the microscopic physics. In this paper, we directly address these challenges by mathematically rephrasing the fluctuation observables in terms of expressions that allow for an intuitively simple physical interpretation. In this way we show explicitly how the measured quantities  relate to the microscopic physics governing strangeness production. Furthermore, we argue that the essential information contained in these fluctuation measures, and indeed more detailed insights, are already captured by measurements of balance functions. We demonstrate this argument through explicit comparisons between rephrased theoretical expressions, full model calculations from PYTHIA and Thermal-FIST, and available experimental data from the ALICE experiment.

\section{Global correlation measures}
\label{sec:global}
The base quantities for constructing the cumulant of net-multistrange baryon number and the Pearson correlation coefficient, are the moments:
\begin{eqnarray}
	&\kappa_{1}(A) = \langle n_A \rangle, \\
	&\kappa_{2}(A) =  \langle n^2_A\rangle - \langle n_A \rangle^2,\\
	&\kappa_{11}(A,B) =  \langle n_A n_B \rangle - \langle n_A\rangle \langle n_B \rangle, \label{eq:kappa11}
\end{eqnarray}
where $A$ and $B$ indicate particle species, $n_A$ the per-event number of particles of species $A$, and $\langle n_A \rangle$ the event-averaged number of particles of species $A$. We furthermore define:
\begin{equation}
	\Delta A = n_{A^+} - n_{A^-}
\end{equation}
as a short-hand for the net-particle number of species $A$, where the subscript indicates electric charge. 

\subsection{Net-$\Xi$ normalised second-order cumulant}

We first study the observable:
\begin{equation}
    \label{eq:k2overk1}
  \frac{\kappa_{2}(\Delta\Xi)}{\kappa_{1}(\Sigma\Xi)} 
  = \frac{\langle
\Delta\Xi^2 \rangle -  \langle \Delta\Xi \rangle^2}{\langle \NXIP \rangle + \langle \NXIM \rangle},
\end{equation}
where:
\begin{equation}
  \Delta\Xi = \NXIP - \NXIM \text{  and  } \Sigma\Xi = \NXIP + \NXIM.
\end{equation}

To gain an intuitive understanding of this observable, we consider the underlying QCD processes driving baryon correlations. In QCD, quarks are always pair-produced and so we expect that any real correlation is introduced by this microscopic pair-production. When we produce a \XIM, the microscopic correlations will be with the antiquarks that were pair-produced in the process. Let us imagine two extreme types of possible QCD physics. If $\XIM$ and $\XIP$ are always produced as pairs, then $\Delta\Xi = 0$ for all events and so $\kappa_{2}(\Delta\Xi)/\kappa_{1}(\Sigma\Xi) = 0$. If on the other hand, $\XIM$ and $\XIP$ are never produced as pairs, meaning that when a $\XIM$ is produced the baryon number is balanced by something else, e.g., an antiproton, and vice verse for $\XIP$, then the production of \XIM and \XIP will be uncorrelated and it is easy to show that for Poisson statistics $\kappa_{2}(\Delta\Xi)/\kappa_{1}(\Sigma\Xi) = 1$. The net-$\Xi$ normalised second-order cumulant therefore measures to what degree $\XIM$ and $\XIP$ are microscopically balanced and is sensitive to both the the pair-production probability, but also to the pair-production correlation length, because a smaller detector will not be able to detect all pairs. 

Recently ALICE has measured $\Xi$-triggered balance functions in \sppt{13} pp collisions~\cite{ALICE:2023asw} for the species most relevant for the observables studied here, namely $\Xi$ and charged kaons. The balance function $B_{\Xi,X}$ is defined as:
\begin{equation}
\label{eq:balanceXi}
\begin{split}
    B_{\Xi,X}(\Delta y) \equiv \frac{1}{N_{\XIM} + N_{\XIP}} \Bigg( & \frac{dN_{\XIM{X}}}{d\Delta y}(\Delta y) - \frac{dN_{\XIM\bar{X}}}{d\Delta y}(\Delta y) \\ 
    & + \frac{dN_{\XIP\bar{X}}}{d\Delta y}(\Delta y) - \frac{dN_{\XIP{X}}}{d\Delta y}(\Delta y) \Bigg),
 \end{split}
\end{equation}
where the first (second) species listed in the $dN_{ij}$ is the trigger (associated) particle and $N_{\XIM} + N_{\XIP}$ is the total number of triggers. The balance function quantifies, on a per-trigger basis, the conditional probability of observing an associated particle with opposite quantum numbers ($X$), corrected for combinatorial background by subtracting same-sign pairs. This ensures that only true correlation signals remain. It has a simple meaning because if one was to measure the balance function with all species and all quantum numbers were conserved then the balance function integrated over all phase space would exactly integrate to the given quantum numbers for $\Xi$: $|Q|=1, |S|=2, |B|=1$ (one has to weight the balance function with the quantum numbers, see~\cite{ALICE:2023asw}). Furthermore, the \emph{measured} balance functions have turned out to vary little with multiplicity~\cite{ALICE:2023asw}, which makes the physics picture simple: one produces more or less particles but the underlying microscopic hadronization mechanism remains the same.

Inspired by Eq.~\ref{eq:balanceXi}, we define the balance number \kxi:
\begin{equation}
\label{eq:kxi}
\begin{split}
\kxi & \equiv \frac{1}{{\langle \NXIM \rangle} + {\langle \NXIP \rangle}}\big( \langle \NXIM\NXIP \rangle - \langle \NXIM(\NXIM - 1) \rangle + \langle \NXIP\NXIM \rangle - \langle \NXIP(\NXIP - 1) \rangle \big) \\
  & \approx \frac{\langle \NXIM\NXIP \rangle - \langle \NXIM(\NXIM - 1)
    \rangle}{\langle \NXIM \rangle},
\end{split}
\end{equation}
where the last equality sign only applies when matter and antimatter are produced in equal amounts as is the case here. In simple terms, $\kxi$ measures how often $\Xi$ baryons are produced in correlated particle-antiparticle pairs, as opposed to being produced independently or with balancing quantum numbers carried by other particles (such as antiprotons). A value of $\kxi$ close to 1 indicates strong local pair production of $\XIP\XIM$, while a value near 0 suggests that $\Xi$ baryons are typically balanced by other antibaryons elsewhere in phase space. This makes $\kxi$ a direct probe of the microscopic hadronization dynamics and correlation length.

Now it is easy to see that we can rewrite Eq.~\ref{eq:k2overk1} as:
\begin{equation}
\label{eq:k2overk1:approx} 
\begin{split}
  \frac{\kappa_{2}(\Delta\Xi)}{\kappa_{1}(\NXIP + \NXIM)}
  & = (1-\kxi) - \frac{\langle \Delta\Xi \rangle^2}{\langle \NXIP \rangle + \langle \NXIM \rangle} \\
  & \approx 1 - \kxi,
\end{split}
\end{equation}
where the last approximation is exact when matter and antimatter are produced in equal amounts. This expression has the two advantages that this form explicitly quantifies the microscopic physics through \kxi \emph{and} that it allows us to compare and contrast the two sets of measurements done by ALICE.

\subsubsection{Model comparison}
\label{sec_model1}

Having obtained a simple relation between the two observables measured by ALICE, we want to compare them with the same models cited in the original paper~\cite{ALICE:2024rnr}. The two models, the Thermal-FIST model and the PYTHIA event generator, have fundamentally different approaches to hadronization and particle production. They represent two contrasting paradigms: thermal statistical hadronization versus perturbative QCD-inspired string fragmentation.

Thermal-FIST (Fast Implementation of Statistical Thermodynamics) \cite{Vovchenko:2019pjl} is a statistical hadronization model designed to describe particle production under the assumption of thermal equilibrium. It implements a hadron resonance gas framework, where particle yields and fluctuations are governed by very few thermodynamic parameters, temperature, chemical potentials, and correlation volumes, and QCD only enters via the spectrum of hadrons that can be produced. In the context of this work, Thermal-FIST is used to simulate event-by-event fluctuations based on global conservation laws and statistical distributions, without modeling the full, dynamical evolution of collisions. Importantly, Thermal-FIST lacks the momentum-space correlations observed in pp collisions, particularly in azimuthal angle and \pT, as it does not simulate particle production mechanisms but rather samples from equilibrium distributions. Therefore the particles produced by the statistical thermal model are boosted using a Blast-Wave model that has been tuned to match the ALICE results for pp collisions at \sppt{13}, which ensures that the \emph{shape} of the \pT spectra are comparable to those measured by ALICE. The correlation length in Thermal-FIST is controlled via the choice of correlation volume, $V_c$, with smaller volumes leading to stronger local conservation effects. In the following we will show two model variations of Thermal-FIST, with two different values for $V_c$: a variation with $V_c = dV/dy$, meaning that the correlation volume is the one expected for a thermal distribution, and a variation with a three times broader correlation volume, $V_c = 3dV/dy$, which was found to be needed to describe the ALICE strangeness enhancement~\cite{Vovchenko:2019kes} and used in Ref.~\cite{ALICE:2024rnr}. In the calculations presented here we have used the exact same procedure as in the ALICE paper and the same version of the model (v1.3.1).

PYTHIA~\cite{Bierlich:2014xba}, in contrast, is a General Purpose Monte Carlo event generator rooted in perturbative QCD for hard processes, combined with the Lund string model for hadronization. In this framework, color fields between partons are represented as strings, which fragment into hadrons through successive string breakings. Strangeness production in PYTHIA arises from the string breaking process, where $s\bar{s}$ pairs are produced with a suppressed but tunable probability relative to up and down quarks. PYTHIA can be extended with rope hadronization~\cite{Bierlich:2014xba}, where overlapping strings form color ropes, enhancing strangeness and baryon production due to the dynamical creation of colour multiplets, giving rise to higher effective string tensions and junction topologies. In the following, we will denote PYTHIA without ropes ``Monash'', after the Monash default tune~\cite{Skands:2014pea}, and ``Ropes'' the tune with ropes.

Finally, we note that in the analysis below, we have employed the same kinematical cuts as in the original papers.\\

\begin{figure}
\centering
\begin{minipage}{0.45\linewidth}
    \includegraphics[width=\linewidth]{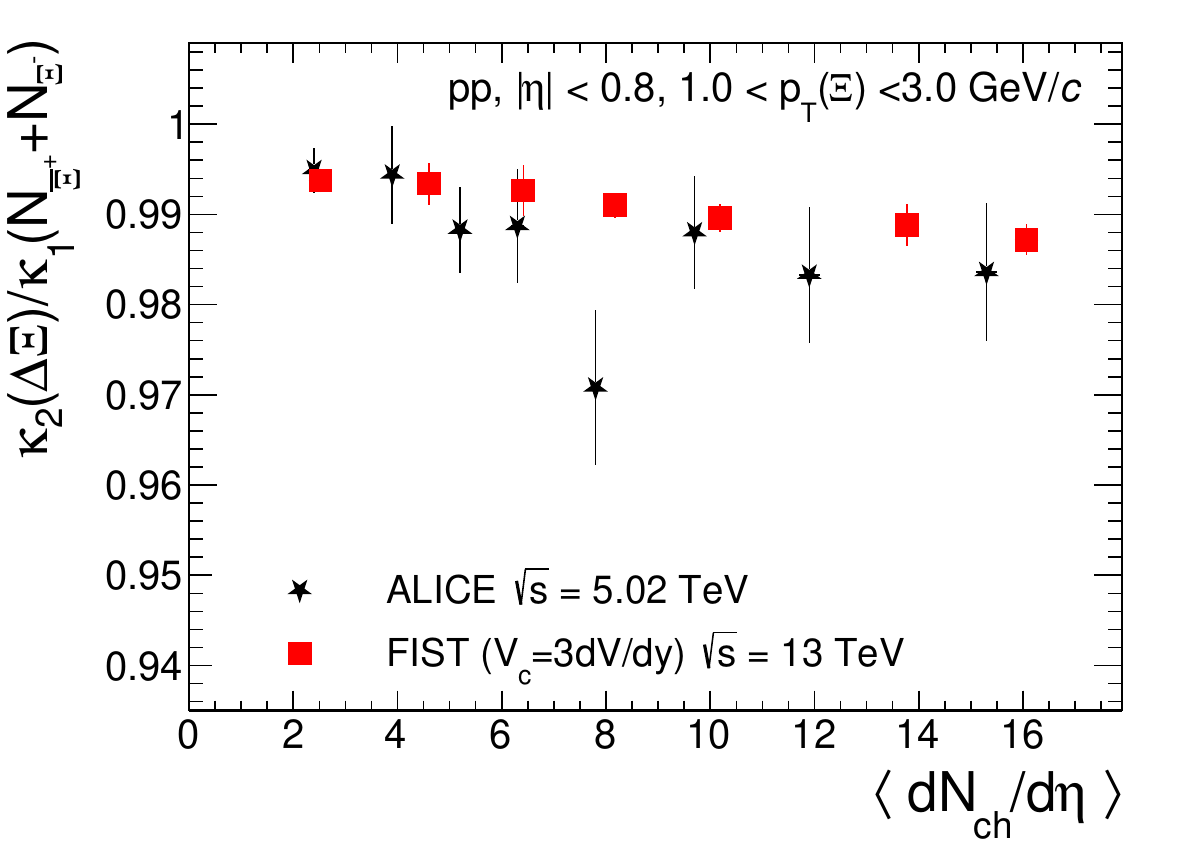}
\end{minipage}
\begin{minipage}{0.45\linewidth}
    \includegraphics[width=\linewidth]{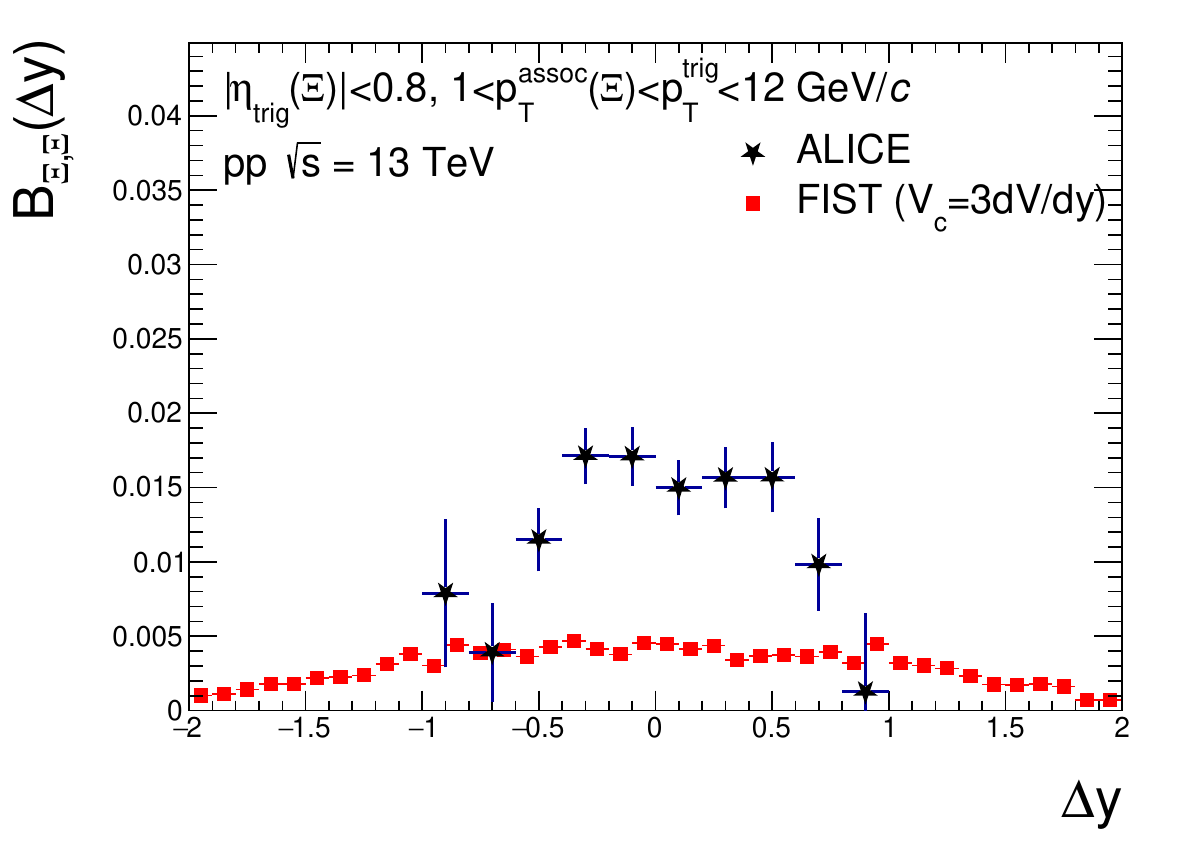}
\end{minipage}

\begin{minipage}{0.45\linewidth}
    \includegraphics[width=\linewidth]{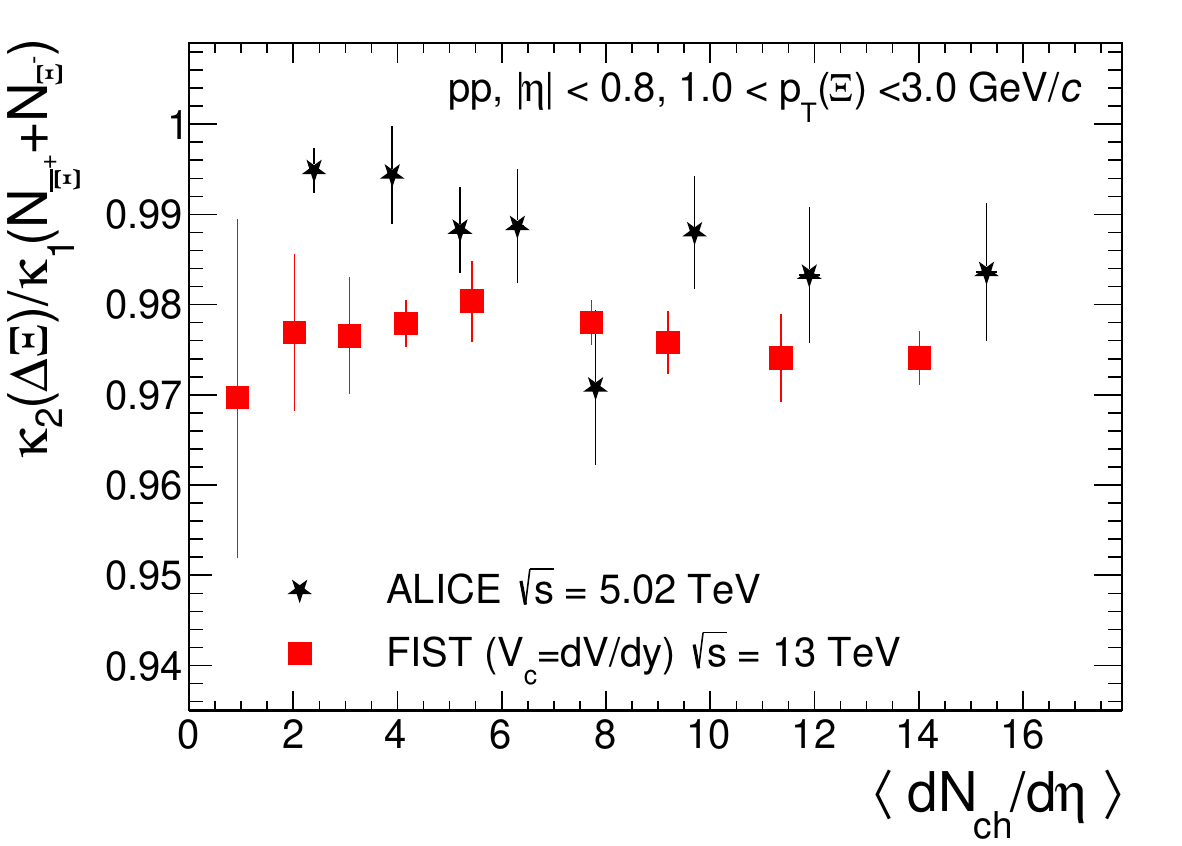}
\end{minipage}
\begin{minipage}{0.45\linewidth}
    \includegraphics[width=\linewidth]{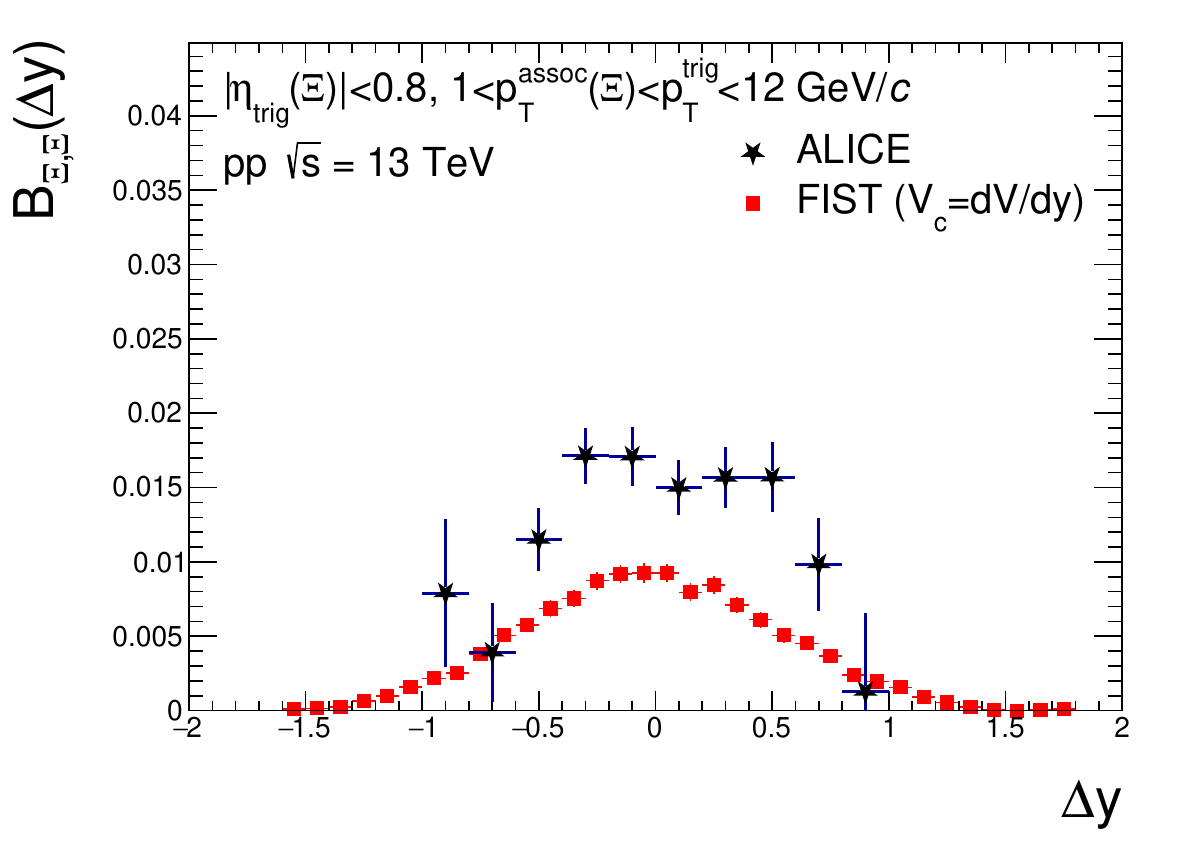}
\end{minipage}

\begin{minipage}{0.45\linewidth}
    \includegraphics[width=\linewidth]{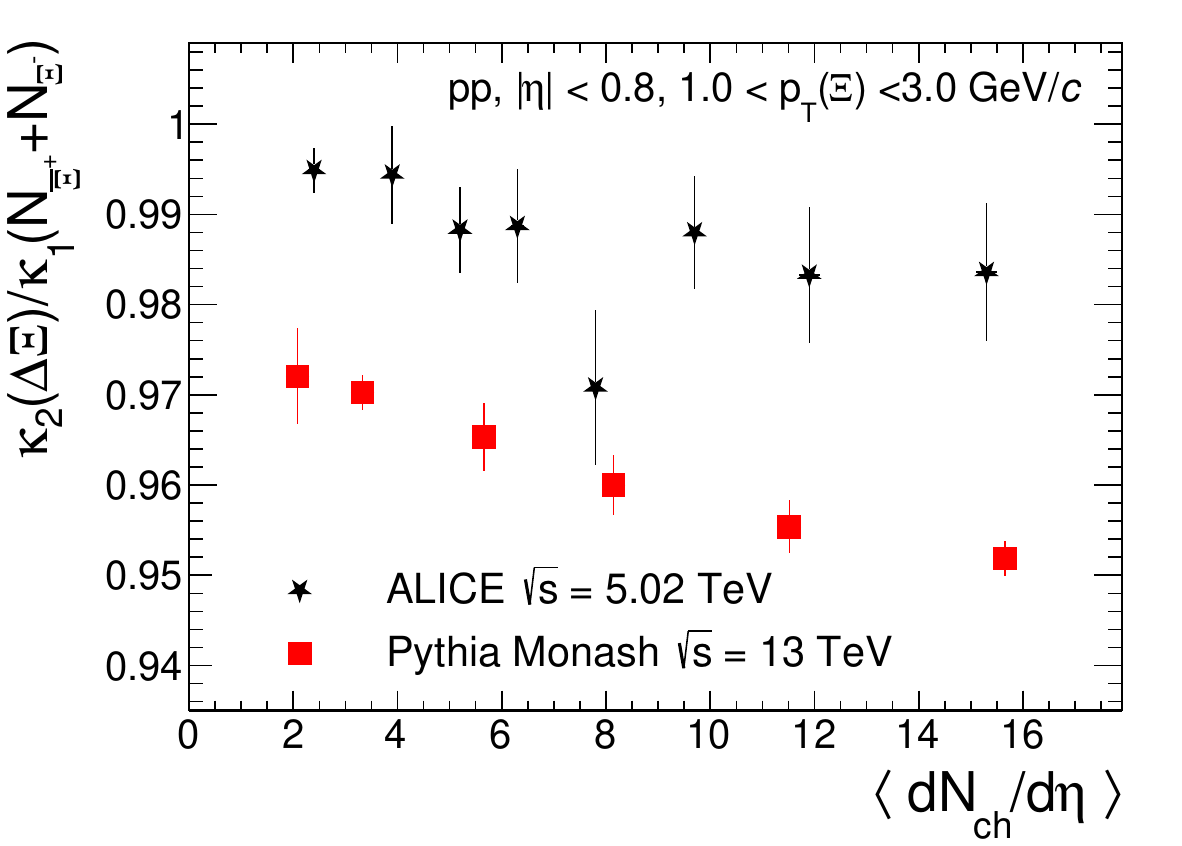}
\end{minipage}
\begin{minipage}{0.45\linewidth}
    \includegraphics[width=\linewidth]{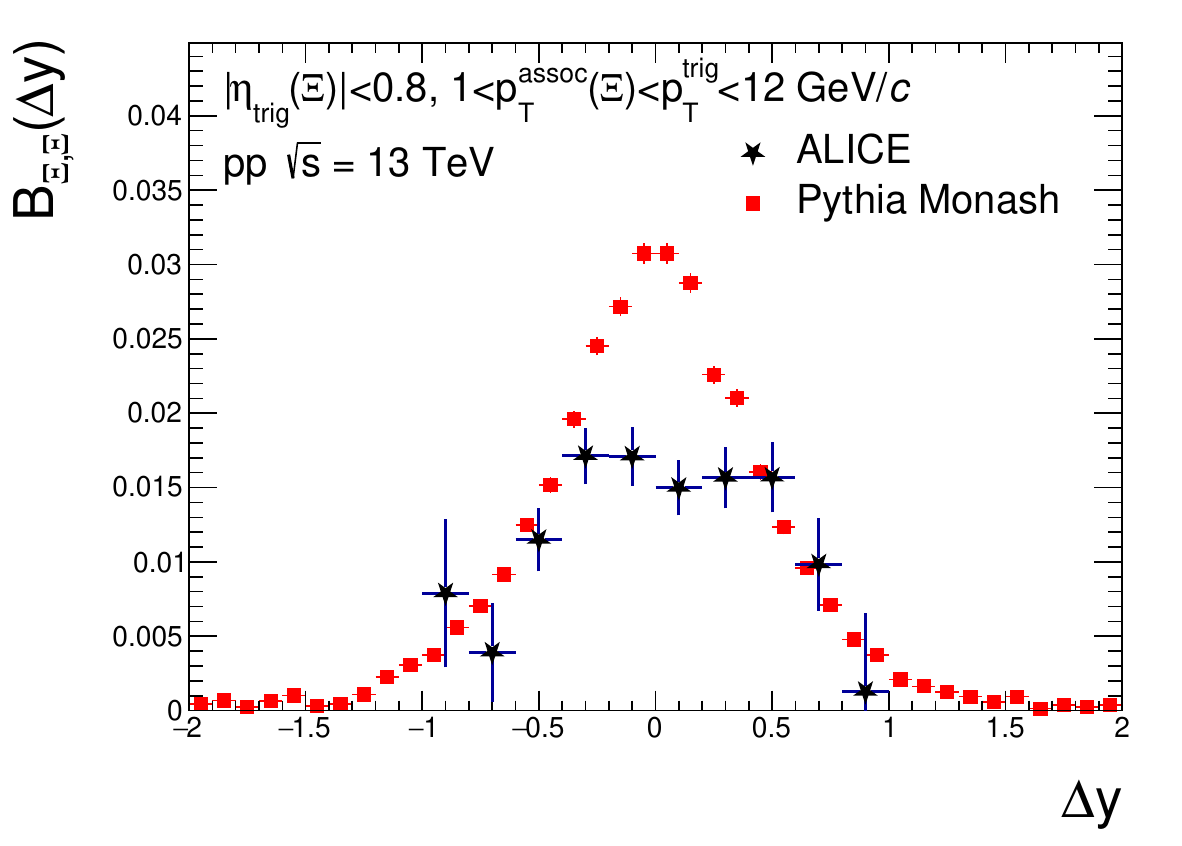}
\end{minipage}

\begin{minipage}{0.45\linewidth}
    \includegraphics[width=\linewidth]{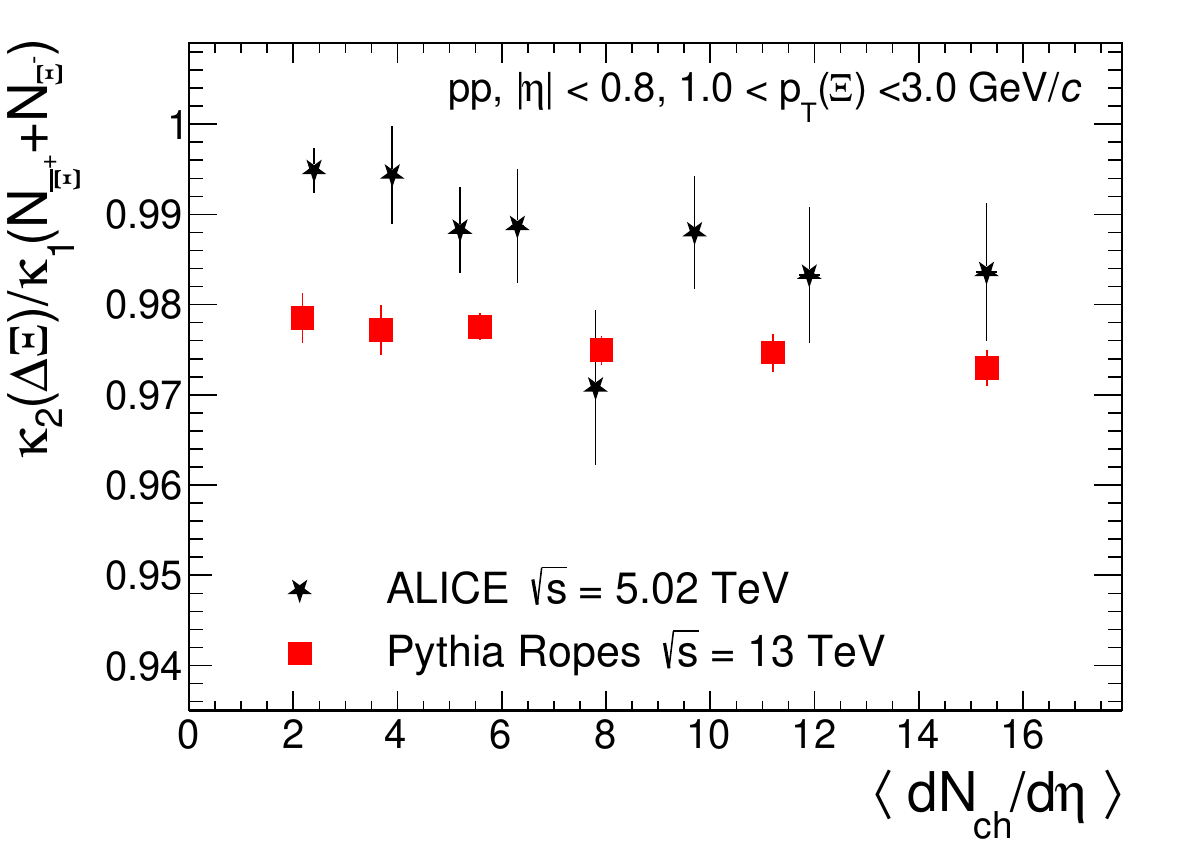}
\end{minipage}
\begin{minipage}{0.45\linewidth}
    \includegraphics[width=\linewidth]{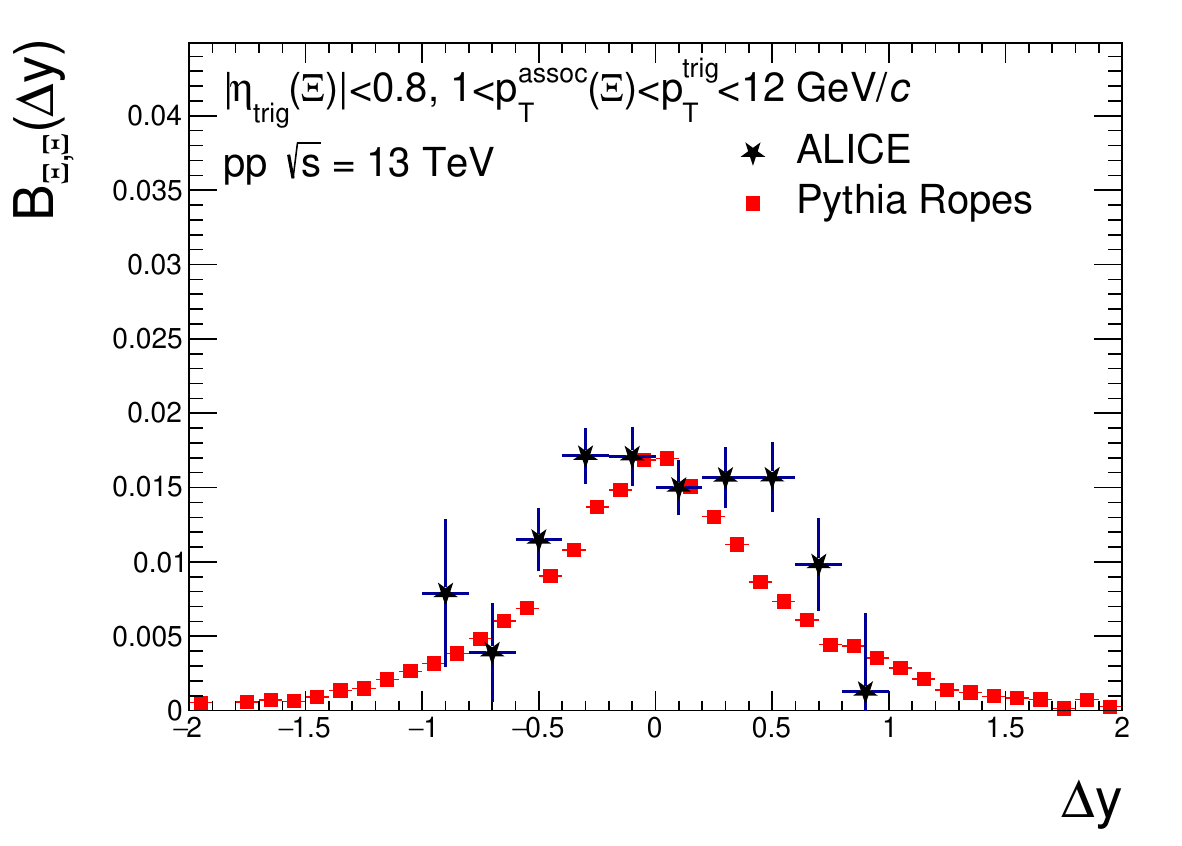}
\end{minipage}
\caption{\label{fig:k2overk1} Left: comparison between the full expression and the approximation given by \eqref{eq:k2overk1} for all models. Right: comparison with the full balance function from Ref.~\cite{ALICE:2023asw} (only statistical uncertainties are shown).}
\end{figure}

Figure~\ref{fig:k2overk1} left shows the comparison of models with the ALICE measurements of $\kappa_{2}(\Delta\Xi)/\kappa_{1}(\Sigma\Xi)$ (similar to Fig.~1 left in Ref.~\cite{ALICE:2024rnr}). This observable is, as discussed above, directly related to the balance function $B_{\Xi,\Xi}$ that has been measured by ALICE in \sppt{13} pp collisions~\cite{ALICE:2023asw} as a function of both azimuthal angle and rapidity separation. While the balance function can be studied as a function of both rapidity ($\Delta y$) and azimuthal angle ($\Delta \phi$), we focus here on the rapidity dependence since Thermal-FIST does not generate realistic azimuthal correlations due to its lack of dynamical particle production mechanisms. We compare with the rapidity separation, by integrating over the azimuthal separation, which has already been done in \eqref{eq:balanceXi}.\\ 

Figure~\ref{fig:k2overk1} right shows a comparison of the same models to the balance function measured in Ref.~\cite{ALICE:2023asw}\footnote{Note that kinematic cuts are different in the two analyses as written in each figure. We have reproduced the analyses faithfully.} In all cases we can see that there is a clear correspondence between the cumulant-ratio in the left figure and the balance function in the right figure. The lower $\kappa_{2}(\Delta\Xi)/\kappa_{1}(\Sigma\Xi)$ is, the larger the integral of the balance function. \\

We emphasize two physics effects, that stand out, when one puts the two comparisons side-by-side as in Fig.~\ref{fig:k2overk1}. \\ Starting with Thermal-FIST, we observe that when the correlation volume is reduced from $3dV/dy$ to $dV/dy$, the balance function narrows. This results in more pairs being balanced inside the detector acceptance, which increases $\kxi$.  This highlights the sensitivity of $\kappa_{2}(\Delta\Xi)/\kappa_{1}(\Sigma\Xi)$ to the correlation length. \\ In PYTHIA, baryons at mid-rapidity are mainly produced by string breakings. Monash, the default PYTHIA tune, only produces baryons though direct diquark breaks, forming $q_1q_2\bar{q_1}\bar{q_2}$. A baryon-antibaryon pair produced by a diquark breaking will share two out of three quarks. Therefore, there is a relatively high probability of pair producing \XIM and \XIP. This probability is expected to be almost independent of multiplicity and so the multiplicity dependence observed for Monash is likely due to the $\langle \pT \rangle$ increase with multiplicity, which means that more $\Xi$ survives the kinematic cuts. With rope hadronization, another production mechanism for baryons is enabled. At high multiplicity, the most common production mechanism for baryons, is by formation and hadronization of junctions~\cite{Bierlich:2014xba}. This type of breaking means that baryons and antibaryons do not have to share any quarks. The difference between Monash and Ropes in Fig.~\ref{fig:k2overk1} highlights the effect of mainly reducing the pair production probability.\\
 
Returning to the comparisons with data in Fig.~\ref{fig:k2overk1}, one surprise is that Thermal-FIST ($3dV/dy$), which does best in describing the ALICE measurement of $\kappa_{2}(\Delta\Xi)/\kappa_{1}(\Sigma\Xi)$ (Fig.~\ref{fig:k2overk1} left), does the worst in describing the balance function (Fig.~\ref{fig:k2overk1} right). There are two effects contributing to this tension. \\
First, the balance functions in Fig.~\ref{fig:k2overk1} right will be dominated by high-multiplicity events because $\Xi$ are much more likely to be produced in high-multiplicity events than at low multiplicity. If we look at the last data point in Fig.~\ref{fig:k2overk1} left, then Thermal-FIST ($3dV/dy$) is at the high end of the error bar while PYTHIA Ropes is at the low end of the error bar. If one calculates the difference in \kxi from the high to the low end, it is a factor ${\approx}2.7$. So by moving the data up and down within the error bar on the left, one can change the extracted balance by almost a factor 3. This means that what appears to be small uncertainties in $\kappa_{2}(\Delta\Xi)/\kappa_{1}(\Sigma\Xi)$, can translate to very large variations for the balance function. 

Secondly, it is important to recognize a fundamental difference in how detector acceptance affects the two observables. The value of $\kxi$, as extracted from the cumulant ratio $\kappa_2(\Delta\Xi)/\kappa_1(\Sigma\Xi)$, reflects the \emph{observed balance within the finite detector acceptance}. In contrast, the balance function shown in Fig.~\ref{fig:k2overk1} (right) is \emph{corrected for detector acceptance effects} using event-mixing techniques, with the aim of extrapolating to what would be measured with full pair acceptance.

This distinction explains why, in the simulations, we do \emph{not} apply a rapidity cut on the associated particles when calculating the balance function. The event-mixing correction accounts for acceptance limitations on the associated side, while still applying cuts on the trigger particle.

To facilitate a direct comparison between the integrated balance function and $\kxi$, we mimic the impact of finite pair acceptance on the balance function by introducing a geometric weighting factor. Specifically, the probability of detecting both particles of a correlated pair decreases linearly with increasing pseudorapidity separation $|\Delta\eta|$. This can be modeled by folding the balance function with a triangular acceptance function, which equals 1 at $\Delta\eta = 0$ and drops to 0 at the detector edges ($|\Delta\eta| = 1.6$).

Applying this folding, the effective $\kxi$ can be approximated by:
\begin{equation}
k_{\Xi} \approx \int_{-1.6}^{1.6} \frac{1.6 - |\Delta\eta|}{1.6} \, B_{\Xi,\Xi}(\Delta\eta) \, d(\Delta\eta).
\end{equation}

This procedure reproduces the value of $\kxi$ to within 1\% accuracy for models with approximately uniform rapidity coverage inside the detector acceptance. This is true both for PYTHIA and Thermal-FIST with $V_c = 3dV/dy$. Notably, deviations arise for Thermal-FIST with $V_c = dV/dy$, where the correlation length becomes comparable to or smaller than the detector acceptance.

This comparison underscores that, once acceptance effects are properly accounted for, the cumulant ratio and the integrated balance function provide equivalent information regarding local baryon number conservation in momentum space. However, unlike the scalar $\kxi$, the balance function retains differential information about the spatial structure of correlations, and is thus a more detailed probe of the underlying dynamics. Finally, it has an intuitive physics meaning as it quantifies microscopic quantum number conservation in momentum phase space.

\subsection{Net-$K$–net-$\Xi$ correlation coefficient}

We now go on to examine the Pearson correlation coefficient $\rho(\Delta \Xi, \Delta K)$. This observable offers a complementary probe of hadronization dynamics to the study of fluctuations of a single particle species, as it is sensitive to local conservation of strangeness, since both $\Xi$ and kaons have strange quarks. It is given by:
\begin{equation}
\label{eq:pearson}
	\rho(\Delta \Xi, \Delta K) = \frac{ \kappa_{11}(\Delta\Xi, {\Delta}K)}{\sqrt{\kappa_2(\Delta \Xi)\kappa_2(\Delta K)}}.
\end{equation}

As before, we aim to express this observable in terms of balance probabilities, isolating the contributions from genuine microscopic correlations driven by shared $s\bar{s}$ quark pairs.
We can reuse \eqref{eq:k2overk1:approx} to approximate: $\kappa_{2}(\Delta\Xi) \approx 2 (1-\kxi) \langle \NXIM \rangle$. But now it is also clear that if we define:
\begin{equation}
\label{eq:kk}
\begin{split}
\kk & \equiv \frac{1}{{\langle \NKM \rangle} + {\langle \NKP \rangle}}\big( \langle \NKM\NKP \rangle - \langle \NKM(\NKM - 1) \rangle + \langle \NKP\NKM \rangle - \langle \NKP(\NKP - 1) \rangle \big) \\
  & \approx \frac{\langle \NKM\NKP \rangle - \langle \NKM(\NKM - 1)
    \rangle}{\langle \NKM \rangle},
\end{split}
\end{equation}
we can approximate: $\kappa_{2}(\Delta K) \approx 2 (1-\kk) \langle \NKM \rangle$. Note that while \kxi is a very small number, the same is not true for \kk since one has for example: $\phi \rightarrow K^+K^-$. 

Looking at the numerator of \eqref{eq:pearson}, we note that the only shared quark content between $\Xi$ baryons and kaons arises from $s\bar{s}$ pairs. Therefore, we expect microscopic correlations to occur predominantly between $\XIM$ and $\KP$, and similarly between $\XIP$ and $\KM$. These correlations contribute negatively to $\rho$, since in both cases $\Delta\Xi\Delta K < 0$. We therefore define the balance factor:
\begin{equation}
\begin{split}
  \kxik & \equiv \frac{1}{{\langle \NXIM \rangle} + {\langle \NXIP \rangle}}\big( \langle \NXIM\NKP \rangle - \langle \NXIM\NKM \rangle + \langle \NXIP\NKM \rangle - \langle \NXIP\NKP \rangle \big) \\
  & \approx \frac{\langle \NXIM\NKP \rangle - \langle \NXIM\NKM
    \rangle}{\langle \NXIM \rangle}.
\end{split}
\end{equation}
Using this definition, the numerator can be written as follows:
\begin{equation}
  \kappa_{11}(\Delta\Xi, \Delta K) \approx -2 \kxik \langle \NXIM \rangle.
\end{equation}

Putting everything together, we obtain the simplified expression for the Pearson correlation coefficient:
\begin{eqnarray}
\nonumber
    \rho(\Delta \Xi, \Delta K) & = & \frac{ \kappa_{11}(\Delta\Xi, {\Delta}K)}{\sqrt{\kappa_2(\Delta \Xi)\kappa_2(\Delta K)}}\\
\nonumber
    & \approx & \frac{-2 \kxik \langle \NXIM \rangle}{\sqrt{2(1-\kxi)\langle \NXIM \rangle 2(1-\kk)\langle \NKM \rangle}} \\ 
\label{eq:pearson-approximation}
    & = & -\kxik \sqrt{\frac{\langle \NXIM \rangle}{(1-\kxi)(1-\kk) \langle \NKM \rangle}},
\end{eqnarray}
which is exact in the limit where matter and antimatter are produced in equal amounts.

This final expression clearly shows that $\rho(\Delta \Xi, \Delta K)$ is governed by the microscopic correlation probability $\kxik$, scaled by the relative yields of $\XIM$ and $K^-$, and modulated by the local quantum number conservation reflected in $(1 - k_{\Xi})$ and $(1 - k_K)$.

\subsubsection{Model comparison}


\begin{figure}

\centering
\begin{minipage}{0.45\linewidth}
    \includegraphics[width=\linewidth]{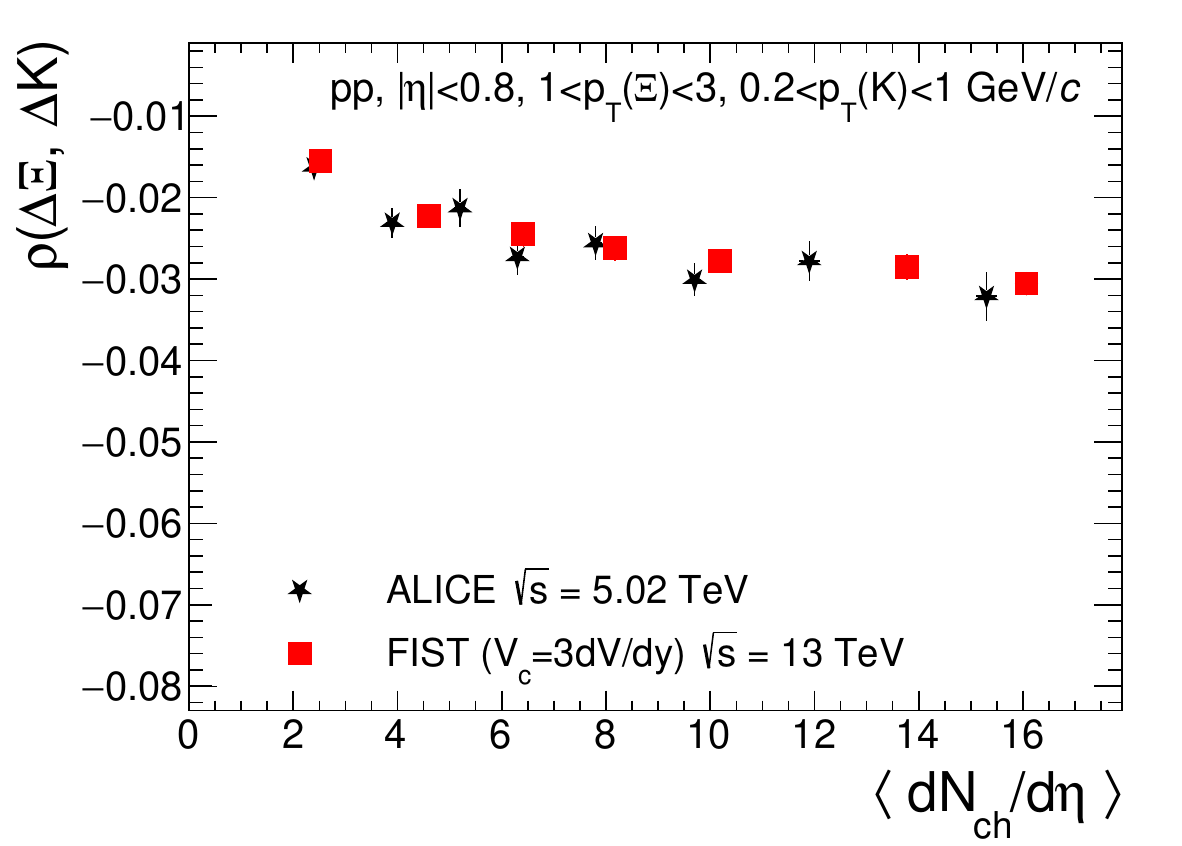}
\end{minipage}
\begin{minipage}{0.45\linewidth}
    \includegraphics[width=\linewidth]{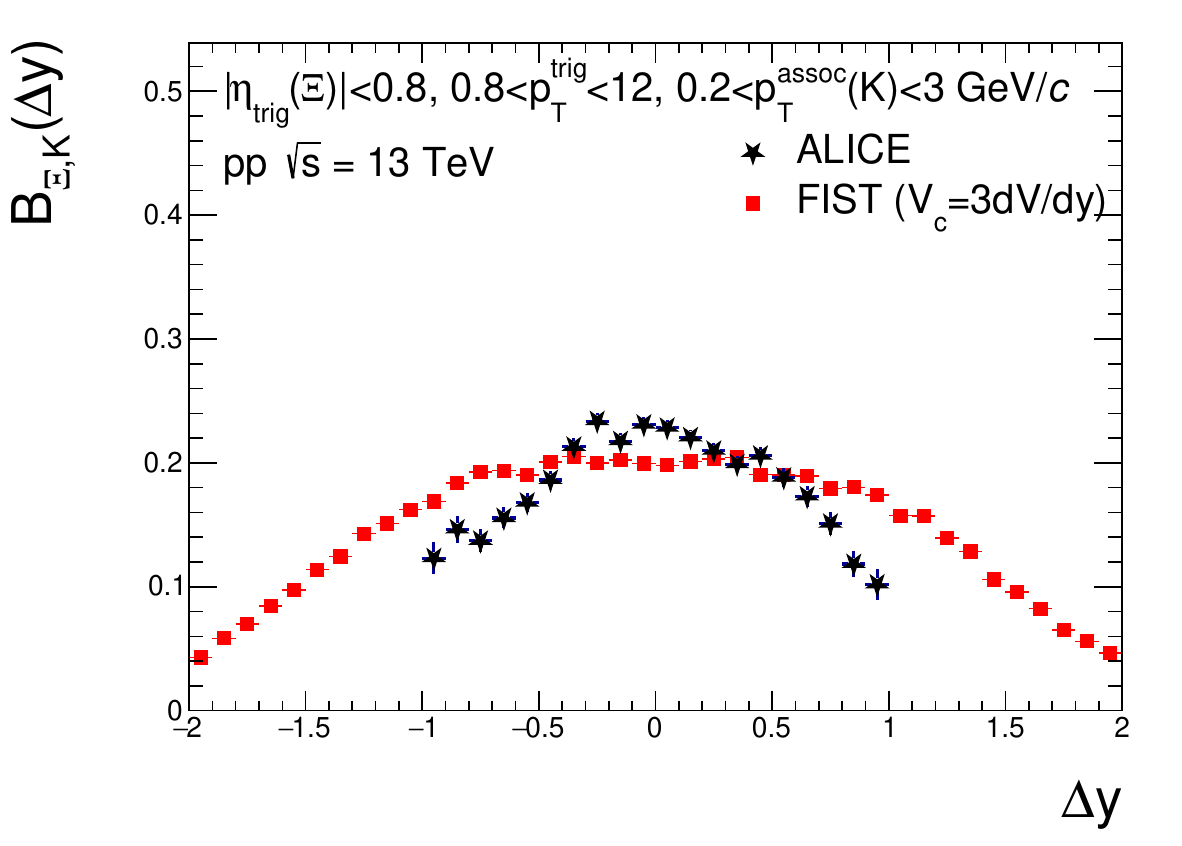}
\end{minipage}

\begin{minipage}{0.45\linewidth}
    \includegraphics[width=\linewidth]{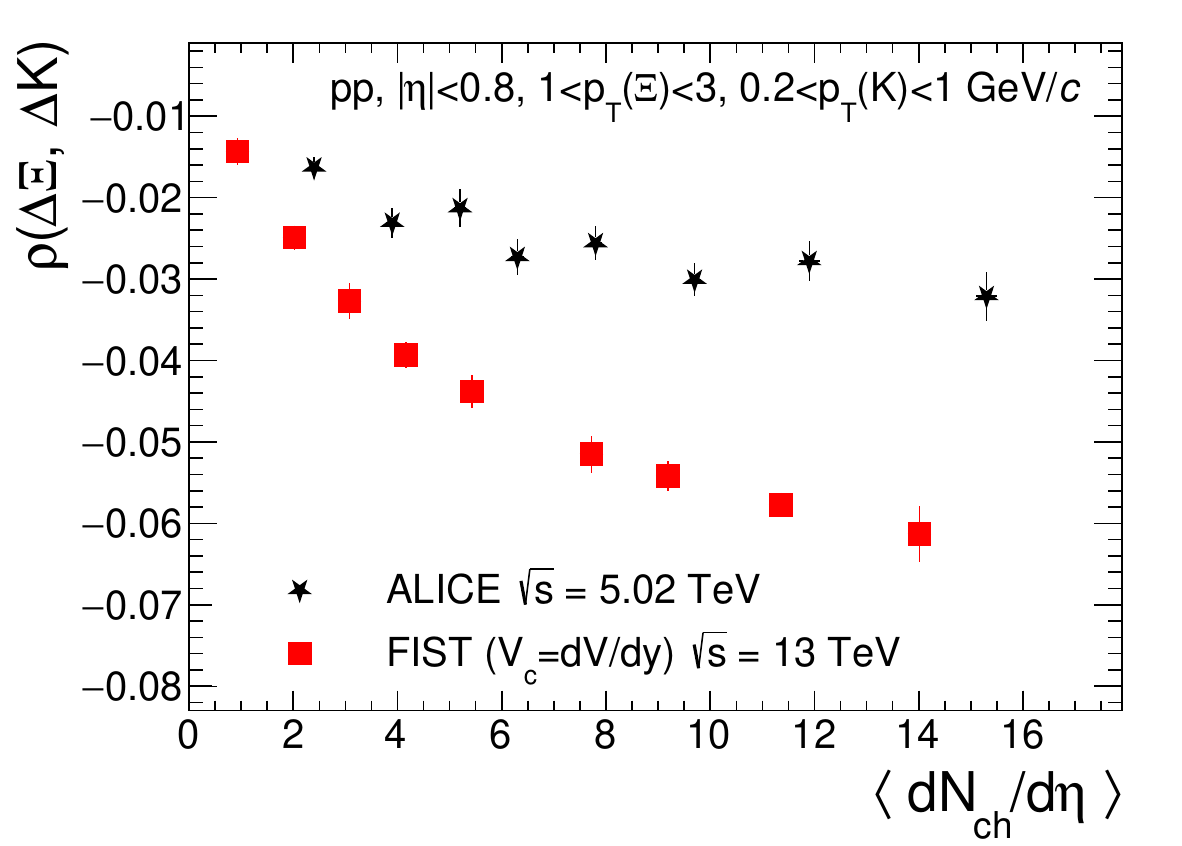}
\end{minipage}
\begin{minipage}{0.45\linewidth}
    \includegraphics[width=\linewidth]{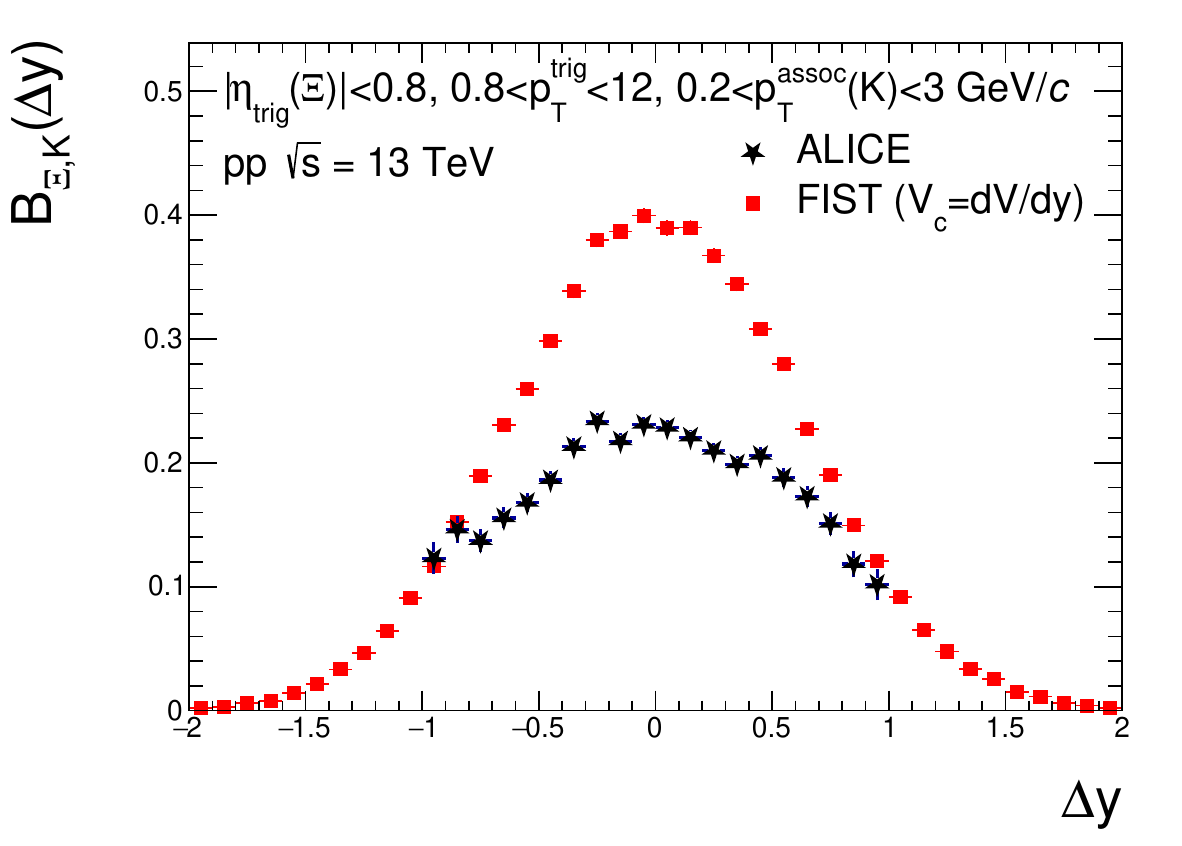}
\end{minipage}

\begin{minipage}{0.45\linewidth}
    \includegraphics[width=\linewidth]{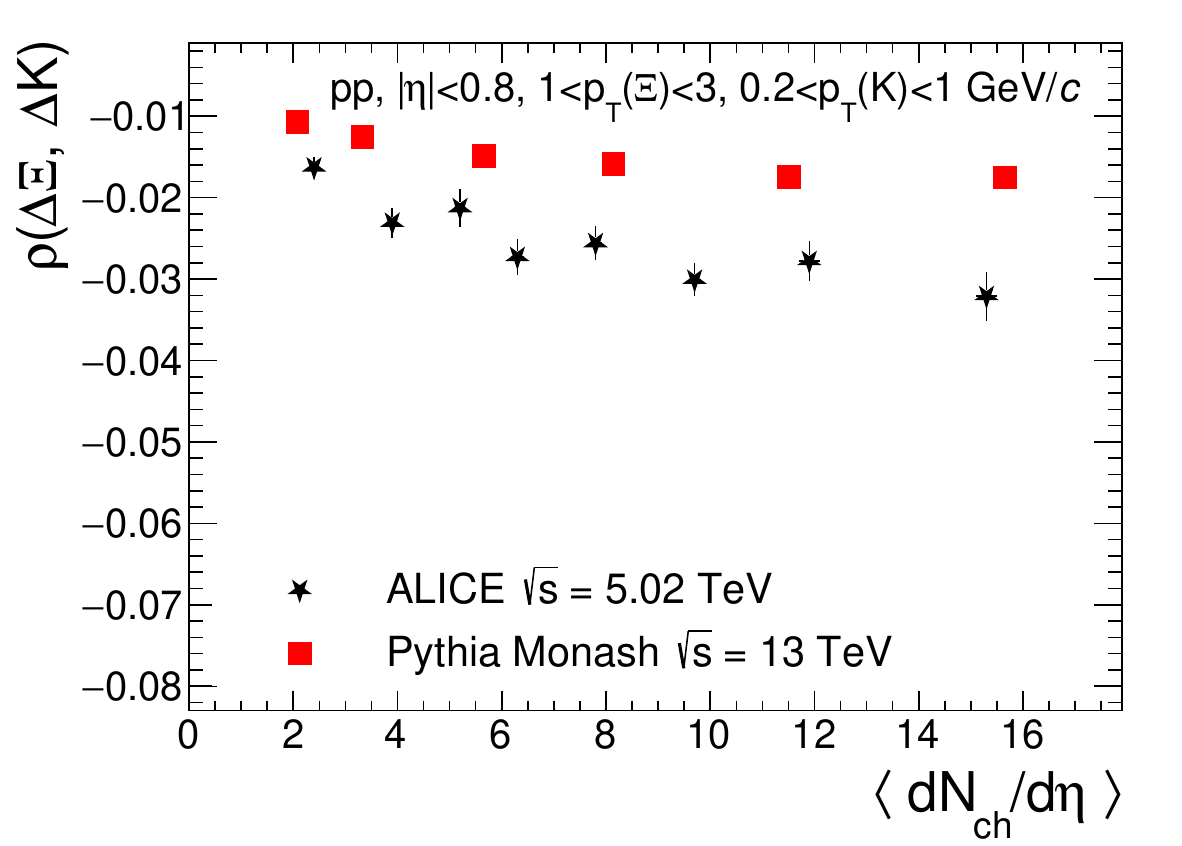}
\end{minipage}
\begin{minipage}{0.45\linewidth}
    \includegraphics[width=\linewidth]{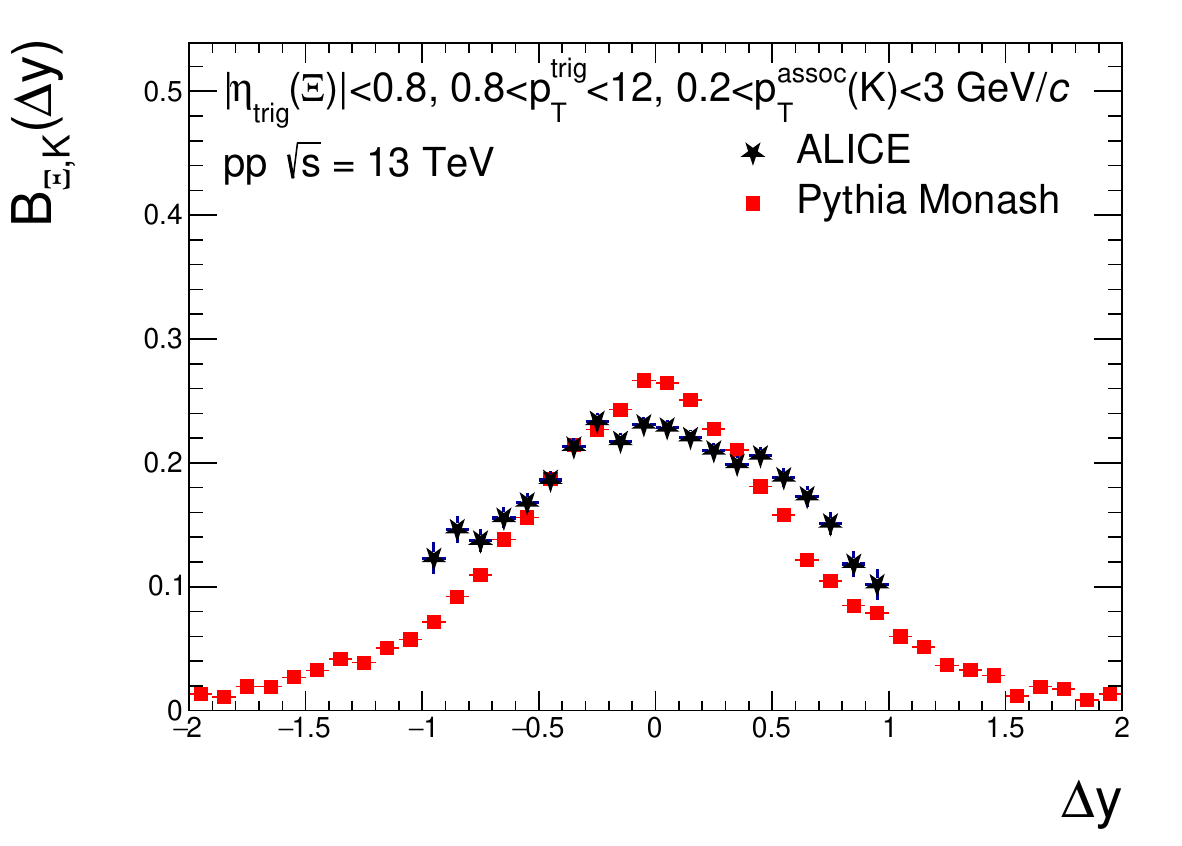}
\end{minipage}

\begin{minipage}{0.45\linewidth}
    \includegraphics[width=\linewidth]{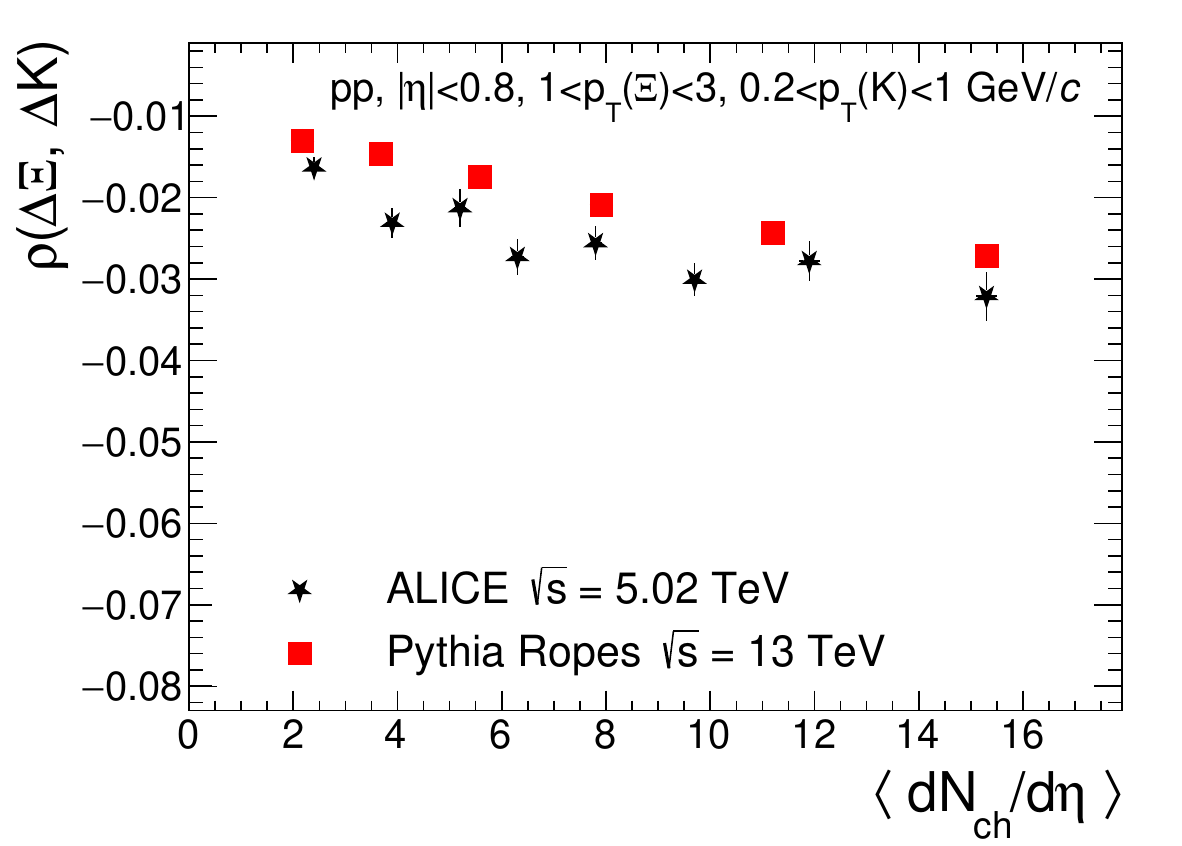}
\end{minipage}
\begin{minipage}{0.45\linewidth}
    \includegraphics[width=\linewidth]{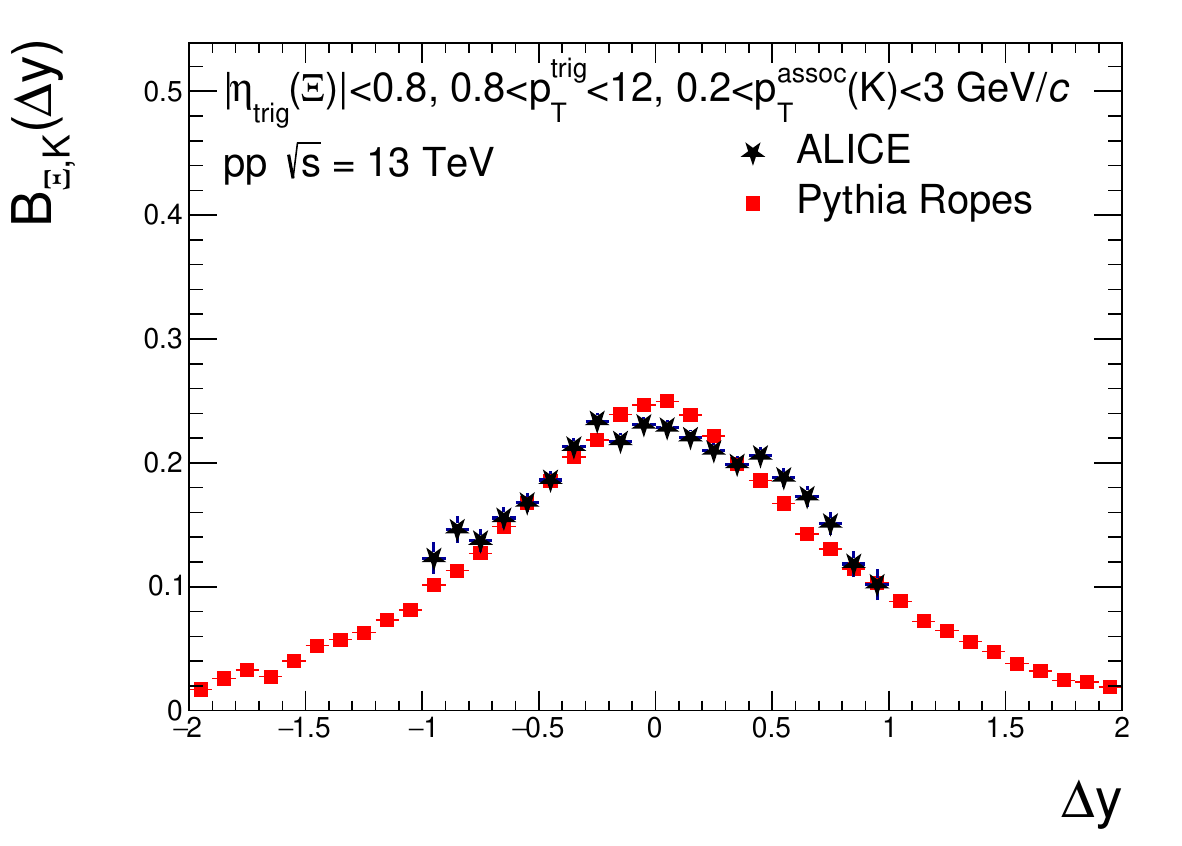}
\end{minipage}
\caption{\label{fig:rho} Model comparisons. Left: comparisons with $\rho(\Delta \Xi, \Delta K)$. Right: comparisons with the full balance function from Ref.~\cite{ALICE:2023asw} (only statistical uncertainties are shown).}
\end{figure}

The balance function $B_{\Xi,K}$, which is related to \kxik, has also been measured by ALICE, and is shown in Fig.~\ref{fig:rho} (right). The balance function is compared with the same model variations as for $\rho$, Fig.~\ref{fig:rho} (left). This balance function directly probes the differential structure of baryon-meson correlations in rapidity space, offering a more detailed view than the scalar quantity, \kxik.

In this comparison, we find that Thermal-FIST with $V_c = 3dV/dy$ reproduces the integral of the balance function well, consistent with its success in describing the magnitude of $\rho$. However, the width of the balance function in Thermal-FIST is significantly broader than observed in data. This highlights once more the lack of realistic dynamic correlations.

Conversely, PYTHIA provides a better description of the shape of the balance function, due to its microscopic string fragmentation dynamics, but struggles with the absolute normalization, likely due to its (known) deficiencies in reproducing strange baryon-to-meson ratios in the measured phase space, possibly due to the lack of radial flow.

These observations highlight that if one is mainly interested in baryon-meson correlations, the balance function is sensitive to both the strength and structure of correlations, i.e., contains richer information than $\rho(\Delta \Xi, \Delta K)$, 


\begin{figure}

\centering
\begin{minipage}{0.45\linewidth}
    \includegraphics[width=\linewidth]{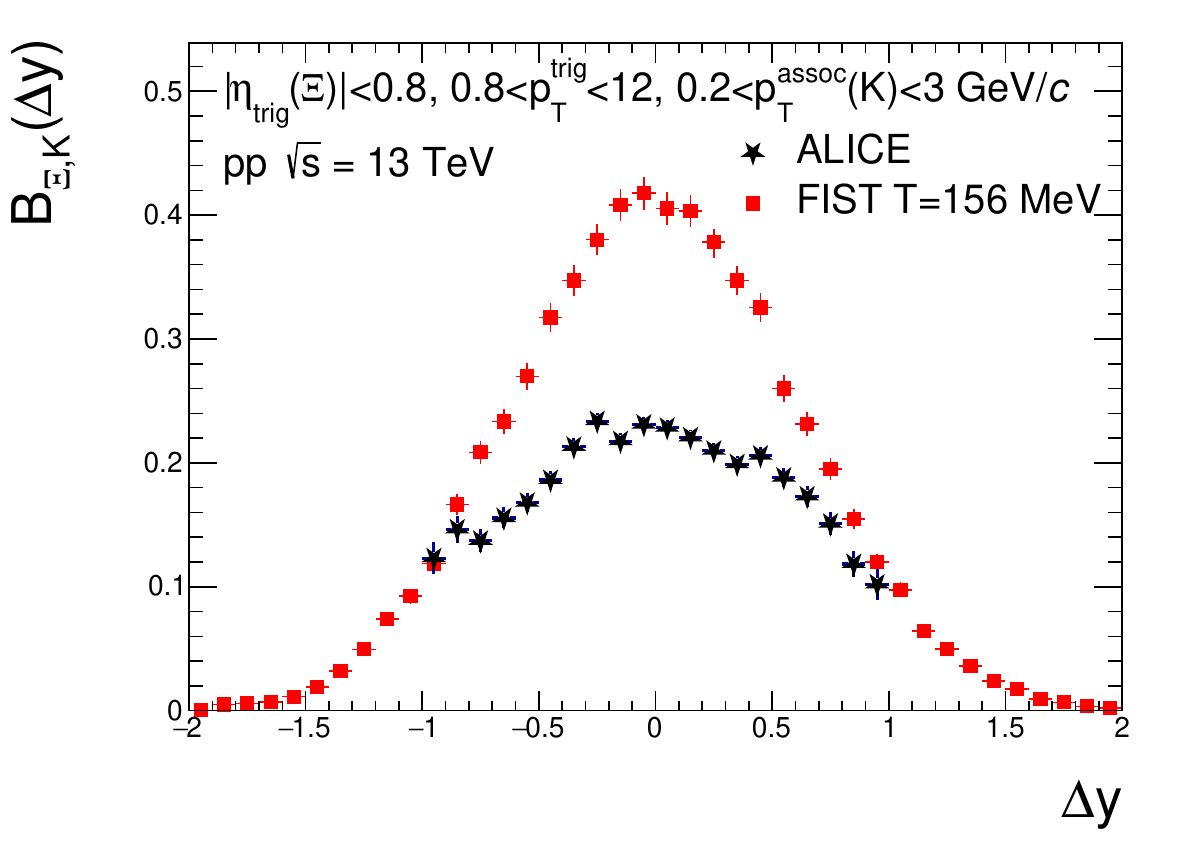}
\end{minipage}
\begin{minipage}{0.45\linewidth}
    \includegraphics[width=\linewidth]{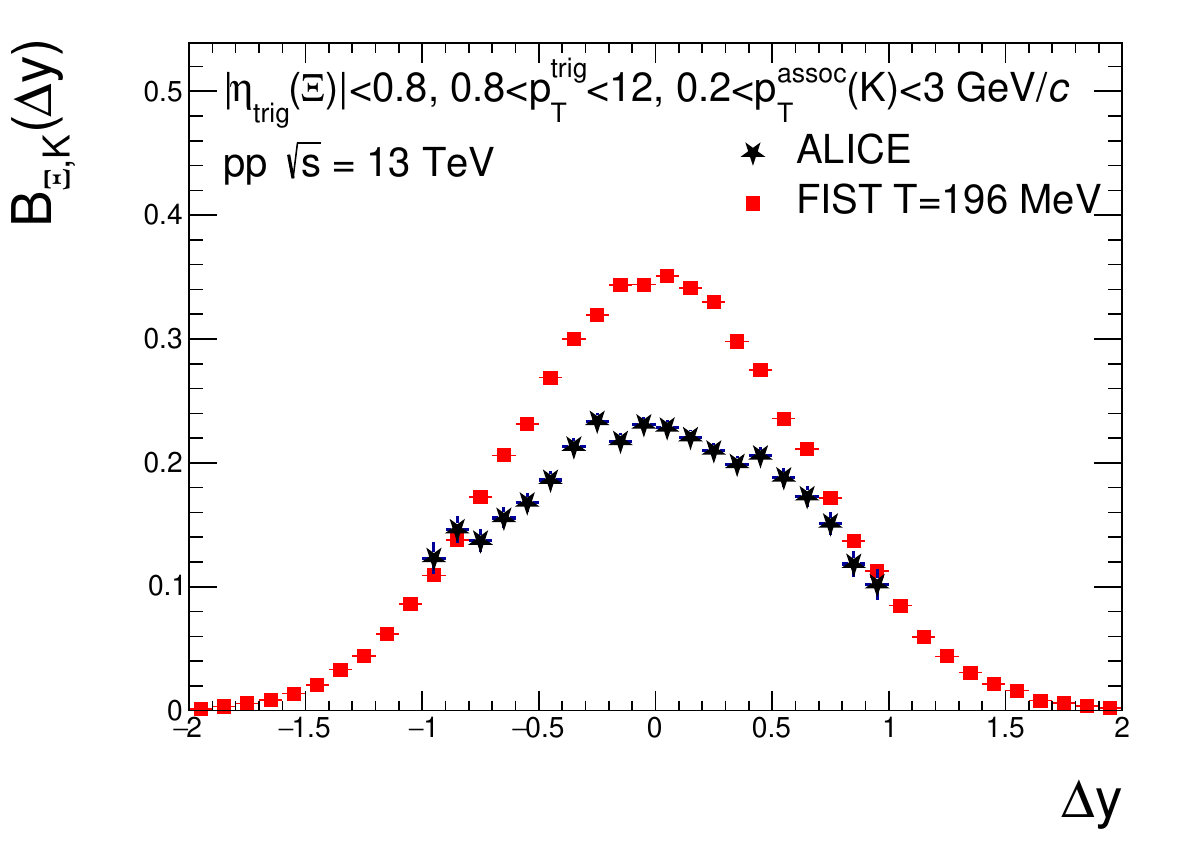}
\end{minipage}

\caption{\label{fig:fist_T} Comparisons with the balance function $B_{\Xi K}$ for Thermal-FIST ($V_c = dV/dy$) with $T = 156\,\text{MeV}$ (left) and $T = 196\,\text{MeV}$ (ŕight).}
\end{figure}

To understand the temperature dependence of the balance functions produced by Thermal-FIST we made two tests where we shifted the temperature up and down by ${\approx}20$ MeV with respect to the default ($T \approx 176\,\text{MeV}$). As can be seen in Fig.~\ref{fig:fist_T} this only affects the balance moderately. If one can understand why the balance only changes moderately with temperature $T$, then this points to that balance functions can play a role in falsifying or validating further statistical thermal models.

\section{Conclusion}
\label{sec:conclusion}

The most central outcome of this study is the surprising realization that both of the recently measured fluctuation observables---the normalized net-$\Xi$ cumulant and the net-$K$–net-$\Xi$ Pearson correlation coefficient---can be directly related to balance properties. This highlights that for systems where matter and antimatter are produced in equal abundances, these fluctuation observables are predominantly governed by local quantum number conservation effects arising from microscopic hadronization dynamics, particularly the sharing of $s\bar{s}$ pairs.

We have demonstrated that, as long as correlations originate from local production processes, global fluctuation measures offer only very limited additional insight beyond what is already encapsulated in differential balance functions. We therefore strongly recommend prioritizing measurements of two-dimensional balance functions in $(\Delta y, \Delta \phi)$, which provide a far more detailed probe of quantum number correlations and thus the underlying hadronization mechanisms.

Even if one would be restricted only to scalar observables such as $\rho(\Delta \Xi, \Delta K)$, using \eqref{eq:pearson-approximation} one can measure the individual contributions to enable a precise identification of where and why models such as PYTHIA and Thermal-FIST succeed or fail in reproducing key features of strangeness production and/or baryon-meson correlations. Given ongoing efforts to improve string-based models, these relations can serve as effective tuning benchmarks.

The poor agreement observed between Thermal-FIST and the measured balance functions further underscores a critical limitation: the absence of a fully dynamical statistical hadronization generator capable of realistically modeling momentum-space correlations. Without such a tool, it remains challenging to assess the microscopic validity of thermal models in small systems, where local correlations and fluctuations play a large role. Finally, we note that balance functions in both small and large systems appears to be very powerful tools for validating the model description of dynamical hadron correlations due to conserved quantum numbers such as strangeness.




\section*{Acknowledgements}

Support from the following research grants are gratefully acknowledged. Vetenskapsrådet contracts 2023-04316 (CB), 2021-05179 (PC), and 2016-05996 (CB).  Knut and Alice Wallenberg foundation contract number 2017.0036 (CB and PC).\\
The authors would also like to thank the ALICE collaboration for providing their code to run the Thermal-FIST and the participants in the ``QCD challenges from pp to AA
collisions'' International Workshop in Münster in September 2024 for inspiring
presentations and discussions.

\bibliographystyle{utcaps}
\bibliography{bibliography}

\providecommand{\href}[2]{#2}\begingroup\raggedright\begin{thebibliography}{10}

\bibitem{ALICE:2022wpn}
{\bfseries ALICE} Collaboration, S.~Acharya {\em et~al.}, ``{The ALICE
  experiment: a journey through QCD},''
  \href{http://dx.doi.org/10.1140/epjc/s10052-024-12935-y}{{\em Eur. Phys. J.
  C} {\bfseries 84} no.~8, (2024) 813},
  \href{http://arxiv.org/abs/2211.04384}{{\ttfamily arXiv:2211.04384
  [nucl-ex]}}.

\bibitem{Christiansen:2024bhe}
P.~Christiansen and P.~Van~Mechelen, ``{Soft QCD Physics at the LHC: highlights
  and opportunities},'' \href{http://arxiv.org/abs/2412.02672}{{\ttfamily
  arXiv:2412.02672 [hep-ex]}}.

\bibitem{Koch:1986ud}
P.~Koch, B.~Müller, and J.~Rafelski, ``{Strangeness in Relativistic Heavy Ion
  Collisions},'' \href{http://dx.doi.org/10.1016/0370-1573(86)90096-7}{{\em
  Phys. Rept.} {\bfseries 142} (1986) 167--262}.

\bibitem{Andronic:2017pug}
A.~Andronic, P.~Braun-Munzinger, K.~Redlich, and J.~Stachel, ``{Decoding the
  phase structure of QCD via particle production at high energy},''
  \href{http://dx.doi.org/10.1038/s41586-018-0491-6}{{\em Nature} {\bfseries
  561} no.~7723, (2018) 321--330},
  \href{http://arxiv.org/abs/1710.09425}{{\ttfamily arXiv:1710.09425
  [nucl-th]}}.

\bibitem{Vovchenko:2019pjl}
V.~Vovchenko and H.~Stoecker, ``{Thermal-FIST: A package for heavy-ion
  collisions and hadronic equation of state},''
  \href{http://dx.doi.org/10.1016/j.cpc.2019.06.024}{{\em Comput. Phys.
  Commun.} {\bfseries 244} (2019) 295--310},
  \href{http://arxiv.org/abs/1901.05249}{{\ttfamily arXiv:1901.05249
  [nucl-th]}}.

\bibitem{Werner:2007bf}
K.~Werner, ``{Core-corona separation in ultra-relativistic heavy ion
  collisions},'' \href{http://dx.doi.org/10.1103/PhysRevLett.98.152301}{{\em
  Phys. Rev. Lett.} {\bfseries 98} (2007) 152301},
  \href{http://arxiv.org/abs/0704.1270}{{\ttfamily arXiv:0704.1270 [nucl-th]}}.

\bibitem{Andersson:1979ij}
B.~Andersson and G.~Gustafson, ``{Semiclassical Models for Gluon Jets and
  Leptoproduction Based on the Massless Relativistic String},''
  \href{http://dx.doi.org/10.1007/BF01577421}{{\em Z. Phys. C} {\bfseries 3}
  (1980) 223}.

\bibitem{Andersson:1983jt}
B.~Andersson, G.~Gustafson, and B.~Söderberg, ``{A General Model for Jet
  Fragmentation},'' \href{http://dx.doi.org/10.1007/BF01407824}{{\em Z. Phys.
  C} {\bfseries 20} (1983) 317}.

\bibitem{Andersson:1983ia}
B.~Andersson, G.~Gustafson, G.~Ingelman, and T.~Sjöstrand, ``{Parton
  Fragmentation and String Dynamics},''
  \href{http://dx.doi.org/10.1016/0370-1573(83)90080-7}{{\em Phys. Rept.}
  {\bfseries 97} (1983) 31--145}.

\bibitem{Bierlich:2024odg}
C.~Bierlich, ``{String Interactions as a Source of Collective Behaviour},''
  \href{http://dx.doi.org/10.3390/universe10010046}{{\em Universe} {\bfseries
  10} no.~1, (2024) 46}, \href{http://arxiv.org/abs/2401.07585}{{\ttfamily
  arXiv:2401.07585 [hep-ph]}}.

\bibitem{Bierlich:2014xba}
C.~Bierlich, G.~Gustafson, L.~L\"onnblad, and A.~Tarasov, ``{Effects of
  Overlapping Strings in pp Collisions},''
  \href{http://dx.doi.org/10.1007/JHEP03(2015)148}{{\em JHEP} {\bfseries 03}
  (2015) 148}, \href{http://arxiv.org/abs/1412.6259}{{\ttfamily arXiv:1412.6259
  [hep-ph]}}.

\bibitem{Bierlich:2022pfr}
C.~Bierlich {\em et~al.}, ``{A comprehensive guide to the physics and usage of
  PYTHIA 8.3},'' \href{http://dx.doi.org/10.21468/SciPostPhysCodeb.8}{{\em
  SciPost Phys. Codeb.} {\bfseries 2022} (2022) 8},
  \href{http://arxiv.org/abs/2203.11601}{{\ttfamily arXiv:2203.11601
  [hep-ph]}}.

\bibitem{Gieseke:2017clv}
S.~Gieseke, P.~Kirchgae\ss{}er, and S.~Pl\"atzer, ``{Baryon production from
  cluster hadronisation},''
  \href{http://dx.doi.org/10.1140/epjc/s10052-018-5585-7}{{\em Eur. Phys. J. C}
  {\bfseries 78} no.~2, (2018) 99},
  \href{http://arxiv.org/abs/1710.10906}{{\ttfamily arXiv:1710.10906
  [hep-ph]}}.

\bibitem{Bellm:2015jjp}
J.~Bellm {\em et~al.}, ``{Herwig 7.0/Herwig++ 3.0 release note},''
  \href{http://dx.doi.org/10.1140/epjc/s10052-016-4018-8}{{\em Eur. Phys. J. C}
  {\bfseries 76} no.~4, (2016) 196},
  \href{http://arxiv.org/abs/1512.01178}{{\ttfamily arXiv:1512.01178
  [hep-ph]}}.

\bibitem{ALICE:2024rnr}
{\bfseries ALICE} Collaboration, S.~Acharya {\em et~al.}, ``{Probing
  Strangeness Hadronization with Event-by-Event Production of Multistrange
  Hadrons},'' \href{http://dx.doi.org/10.1103/PhysRevLett.134.022303}{{\em
  Phys. Rev. Lett.} {\bfseries 134} no.~2, (2025) 022303},
  \href{http://arxiv.org/abs/2405.19890}{{\ttfamily arXiv:2405.19890
  [nucl-ex]}}.

\bibitem{ALICE:2023asw}
{\bfseries ALICE} Collaboration, S.~Acharya {\em et~al.}, ``{Studying
  strangeness and baryon production mechanisms through angular correlations
  between charged \ensuremath{\Xi} baryons and identified hadrons in pp
  collisions at $ \sqrt{s} $ = 13 TeV},''
  \href{http://dx.doi.org/10.1007/JHEP09(2024)102}{{\em JHEP} {\bfseries 09}
  (2024) 102}, \href{http://arxiv.org/abs/2308.16706}{{\ttfamily
  arXiv:2308.16706 [hep-ex]}}.

\bibitem{Vovchenko:2019kes}
V.~Vovchenko, B.~D\"onigus, and H.~Stoecker, ``{Canonical statistical model
  analysis of p-p , p -Pb, and Pb-Pb collisions at energies available at the
  CERN Large Hadron Collider},''
  \href{http://dx.doi.org/10.1103/PhysRevC.100.054906}{{\em Phys. Rev. C}
  {\bfseries 100} no.~5, (2019) 054906},
  \href{http://arxiv.org/abs/1906.03145}{{\ttfamily arXiv:1906.03145
  [hep-ph]}}.

\bibitem{Skands:2014pea}
P.~Skands, S.~Carrazza, and J.~Rojo, ``{Tuning PYTHIA 8.1: the Monash 2013
  Tune},'' \href{http://dx.doi.org/10.1140/epjc/s10052-014-3024-y}{{\em Eur.
  Phys. J. C} {\bfseries 74} no.~8, (2014) 3024},
  \href{http://arxiv.org/abs/1404.5630}{{\ttfamily arXiv:1404.5630 [hep-ph]}}.

\end{thebibliography}\endgroup

\end{document}